# Weak Forms of Monotonicity and Coordination-Freeness

Early Research Report.

Daniel Zinn
daniel.zinn@logicblox.com

LogicBlox, Inc.
1349 W Peachtree St NW
Atlanta, GA 30309 USA

## ABSTRACT

Our earlier work [3] has shown that the classes of queries that can be distributedly computed in a coordination-free manner form a strict hierarchy depending on the assumptions of the model for distributed computations. In this paper, we further characterize these classes by revealing a tight relationship between them and certain weakened forms of monotonicity.

## 1. INTRODUCTION

In [3], we have shown that the classes of coordination-free queries form a proper hierarchy based on various extensions to the network of relational transducer model presented in [1]. In the basic model $\mathcal{N}_0$, input data is arbitrarily distributed amongst the nodes in a distributed system. Furthermore, no node has any information about the distribution process. Ameloot et al. have shown that under this model, the class of coordination-free queries are the monotone queries. In the model $\mathcal{N}_1$, data is partitioned according to a *Policy*, much like a hash- or range-partitioning scheme. That is, for every potential input fact, the policy dictates on which node(s) the fact is stored. Furthermore, network nodes are equipped with local knowledge that allows answering the question "Would the fact $f$ be allocated to me?". The set of queries that can be distributedly computed without coordination in this extened model is strictly larger than the monotone queries. In the model $\mathcal{N}_2$, data is stored with replication. Under the model $\mathcal{N}_2$, even non-stratified queries such as the win-move game can be computed in a coordination-free manner. To distribute data according to the $\mathcal{N}_2$ model, each domain constant is assigned to a host. Ground facts are then distributed to the union of all nodes assigned to the fact's constant (or a fixed node in case the fact is nullary). In the last model $\mathcal{N}_3$, we further equip every node with knowledge about the active domain of the global input. Here, *every query* is coordination-free. Abbreviating the set of monotone queries with $\mathcal{M}$, the set of all computable queries with $\mathcal{C}$, and the sets of coordination-free queries according to model $X$ as $\mathcal{F}[X]$, then the following hierarchy holds as shown in [3]:

$$\mathcal{M} = F[\mathcal{N}_0] \subsetneq \mathcal{F}[\mathcal{N}_1] \subsetneq \mathcal{F}[\mathcal{N}_2] \subsetneq F[\mathcal{N}_3] = \mathcal{C}$$

While [3] only gave examples of queries in $\mathcal{F}[\mathcal{N}_1]$ and $\mathcal{F}[\mathcal{N}_2]$, we here characterize the two classes exactly. It turns out that the coordination-free queries according to model $\mathcal{N}_1$ and $\mathcal{N}_2$ coincide with the queries that fulfill certain weaker notions of monotonicity.

Recall that a query $Q$ is monotone if $Q(I) \subseteq Q(I \cup I')$ for all inputs $I$ and $I'$. Without loss of generality[1], we can restrict our attention to $I'$ that only contain one input fact $f$. A query is thus monotone iff its result is non-shrinking for any fact $f$ that is added to the input. The weaker notions of monotonicity, namely adom-monotonicity and weak-adom-monotonicity, are similar except that restrictions are imposed to the fact $f$ that is added to $I$. An adom-monotone query has a non-shrinking result only for facts that contains *at least one* constant that does not appear in $I$. Further, weak-adom-monotone queries have non-shrinking output for non-nullary facts that contain *only* constants not occurring in $I$.

Denoting the classes of adom-monotone and weak-adom-monotone queries with $\mathcal{M}_{adom}$ and $\mathcal{M}_{weak-adom}$, respectively, our main result shows:

$$\mathcal{F}[\mathcal{N}_1] = \mathcal{M}_{adom} \quad \text{and} \quad \mathcal{F}[\mathcal{N}_2] = \mathcal{M}_{weak-adom}$$

This short report is structured as usual: the preliminaries in Section 2 reiterate or reference basic database concepts, the definitions of relational transducers and transducer networks as well as the definitions for coordination-freeness. Section 3 presents our main results. We then briefly discuss related work and conclude in Section 4.

## 2. PRELIMINARIES

This section briefly introduces basic notations. The section can be skipped if the reader is familiar with the contents of [3].

### 2.1 Database Basics

We fix the domain *dom*, a countable set of constants. A *relational schema* is a set of relation symbols $r_1, r_2, \ldots, r_n$

---

[1]A simple inductive argument shows that the two notions are equivalent.

with associated arities $\alpha(r_i)$. A database *instance* $I$ over a schema $\sigma$ associates an $\alpha(r_i)$-ary relation over the *domain* with each relational symbol $r_i \in \sigma$. When convenient, we identify database instances with sets of *facts* of the form $r_i(a_1, \ldots, a_{\alpha(r_i)})$ in the usual way. Furthermore, the active domain of an instance $adom(I)$ is the set of constants from $dom$ that are used in the facts of $I$.

A query is a mapping between finite database instances. A query $Q$ is *monotone* if $Q(I) \subseteq Q(I \cup I')$ for all database instances $I$ and $I'$.

## 2.2 Transducer Networks and Coordination-Freeness

Transducer networks have been introduced by Ameloot et al. [1]. We here only give an intuitive understanding of transducer networks and refer the reader to [1] or [3] for technical details.

**Transducer networks.** In transducer networks, there is a *relational transducer* located at each node of a connected network graph. A relational transducer is a collection of deterministic queries that are executed in an endless loop. In each step, the query outputs define (1) the incremental *output* of the transducer, (2) the *message* data that is sent to the neighbors according to the network graph, and (3) the *updates* to the local memory of the transducer. The query input is the union of the relational data over four relational schemas: input relations, system relations, local memory relations, and incoming message relations. Input and system relations cannot be modified and are pre-loaded with part of the input data and relevant information about the transducer network, respectively. Memory relations are initially empty and are updated according to the output of the transducer's queries at the end of each step. Incoming message relations are either empty or contain one fact that is consumed by the transducer.

A step (or transition) of the transducer amounts to computing the result of each of these queries and performing memory-updates, transducer output, and message sending. Memory updates are performed according to designated "to-delete" and "to-insert" output relations of the queries. The final output of the transducer is understood as the union of all output performed at any step. That is, output that has been asserted cannot be retracted later on. Messages to neighbors will eventually be delivered and accumulate at neighbor's *input buffers*. Each node is associated an input buffer, which is a bag of facts over the message schema. The network graph is connected such that any node can reach any other node, possibly routing data across intermediary nodes.

Transducer queries are deterministic. The following properties of transducer networks model the inherent non-determinism found in real distributed systems. Even though the network messages computed by a transducer in a step is a set of facts, each transducer can consume network input only one-fact at a time. Here, (1) the order in which facts are consumed (i.e., taken out from the input buffer) is non-deterministically chosen. Furthermore, (2) a transducer might non-deterministically choose to step without consuming any input. Such a step is called a heart-beat transition. Finally, (3) the network steps by non-deterministically choosing one node to step. As usual, there are fairness criteria in place that avoid nonsensical executions: each transducer has to perform infinitely many heartbeat transitions, and every fact that is placed in an input buffer is eventually consumed by the transducer on this node.

A transducer network is defined by the network graph $N$ and the set of queries of a transducer $\mathcal{T}$, formally it is a pair $(N, \mathcal{T})$. Each node in the system deploys the same transducer queries. The transducer network's input is partitioned (with possible replication) across the nodes of the network. The network's output is the union of the output of all transducers in the network.

**Computing a query.** Given an input $I$ and a transducer network $(N, \mathcal{T})$, the output of the transducer network is determined by the way input is distributed, and by the non-deterministic choices made during its execution as explained in (1)–(3) above. We say a transducer $\mathcal{T}$ *distributedly computes* a query $Q$ if for any network $N$, any input distribution, and any non-deterministic choices, the transducer network outputs $Q(I)$. In other words, to compute a query, the transducer network's output must be deterministic despite all internal non-determinism during query execution.

**Extensions.** In our earlier work [3], we extended this basic model (which we call $\mathcal{N}_0$) by allowing nodes to have limited knowledge about the distribution process of the data, resulting in the models $\mathcal{N}_1$, $\mathcal{N}_2$, and $\mathcal{N}_3$.

In short, for models $N_i, i \geq 1$, there exists a function $\mathcal{P}$ which we call the Partitioning Policy. $\mathcal{P}$ assigns each *possible* input fact (i.e., ground fact from the Herbrand Base of the input schema) to a non-empty set of nodes. Actual input data is then distributed according to $\mathcal{P}$. Formally: fix an input schema $\sigma$. $\mathcal{P}$ assigns each fact in the infinite set

$$\{R(c_1, \ldots, c_n) \mid R \in \sigma, \alpha(R) = n, c_i \in dom \text{ for all } i\}$$

to a non-empty set of nodes. Each fact $f$ is then placed into the input relations of the transducer at node(s) $\mathcal{P}(f)$ *if* $f$ is part of the input EDB. Furthermore, each transducer node has access to $\mathcal{P}$ via a decision oracle (part of the systems relations). That is, node $n_j$ can answer the boolean query "Is $n_j \in \mathcal{P}(f)$?" for facts $f = R(c_1, \ldots, c_n)$ that node $n_j$ can construct. A node can construct a fact $f$ if all constants $c_i$ have become known to the node: either via its initial input relations, or during the computation once data has been exchanged.

For $\mathcal{N}_1$ and $\mathcal{N}_3$ transducers, we require that a transducer network always computes the correct result for *any* distribution Policy $\mathcal{P}$. In the model $\mathcal{N}_2$, we restrict our attention to *compatible* distribution policies, which provide a certain kind of data replication. Here, we require that the transducer network computes the correct result for any compatible policy. Specifically, for $\mathcal{N}_2$ transducer networks we require the existence of a function $F : dom \to 2^N \setminus \{\emptyset\}$ assigning each domain constant to a non-empty set of nodes. Then a partitioning policy $\mathcal{P}$ is *compatible* with model $\mathcal{N}_2$ if for all non-nullary relations the following holds:

$$\mathcal{P}(R(c_1, \ldots, c_n)) = \bigcup_{i=1,\ldots,n} F(c_i).$$

In $\mathcal{N}_3$ transducers, we allow each transducer to have access to the active domain of the global input.

More details and formal definitions of transducer networks and the variations thereof are presented in [3] section 5.

**Coordination-Freeness.** Like in our previous work [3], we also follow Ameloot [1] with the definition of coordination-freeness. A transducer $\mathcal{T}$ that distributedly computes a query $Q$ is *coordination-free with respect to model $X$* for $X \in \{\mathcal{N}_0, \mathcal{N}_1, \mathcal{N}_2, \mathcal{N}_3\}$ if for every input $I$ and every network $N$, there exists a distribution policy $\mathcal{P}$ that is compatible with $X$ for which when the transducer network $(N, \mathcal{T})$ is run with only heartbeat transitions, it already produces the correct result.

A query $Q$ is *coordination-free in $X$* if there exists a coordination-free type $X$ transducer $\mathcal{T}$ that computes $Q$. We denote the class of coordination-free queries with respect to model $X$ by $\mathcal{F}[X]$.

## 3. COORDINATION-FREE QUERIES

**Definition 3.1 (Adom-Monotonicity)** Let $I$ be a database instance over a schema $\sigma$, and $R$ a relation symbol in $\sigma$ of arity $n$. A query is *adom-monotone* if $Q(I) \subseteq Q(I \cup \{f\})$ for all inputs $I$ and facts $f = R(c_1, \ldots, c_n)$ that contain at least one constant $c_i$ that does not occur in $I$, i.e., $c_i \notin adom(I)$ for some $i$. We denote the class of adom-monotone queries as $\mathcal{M}_{adom}$.

In a sense, adom-monotone queries allow the same degree of non-monotonicity as semi-positive Datalog. Let $\mathcal{SP}$ denote the set of queries $Q$ for which there exists a semi-positive Datalog program that computes $Q$.

**Lemma 3.2** $\mathcal{SP} \subseteq \mathcal{M}_{adom}$.

PROOF. Fix a query $Q$ over the schema $\sigma$, which contains an $n$-ary relation $S$. Assume towards a contradiction that there is a database instance $I$ over schema $\sigma$ and a fact $f = S(c_1, \ldots, c_n)$ with some $c_i \notin adom(I)$ such that $Q(I) \not\subseteq Q(I \cup \{f\})$.

Let $P$ be the semi-positive Datalog$^\neg$ program computing $Q$. Now construct a positive Datalog program $P'$ from $P$ via the obvious transformation: For each relation symbol $R \in \sigma$, add a new relation $R^c$; and replace each occurrence of $\neg R$ in $P$ by $R^c$. Note, $P'$ is a positive Datalog program and thus monotone. For a database instance $I$ over $\sigma$, let $I^c$ denote its complement with respect to its active domain using "$^c$"-annotated relation symbols:

$$I^c = \{R^c(\bar{a}) \mid R \in \sigma, \alpha(R) = k, \bar{a} \in adom(I)^k, R(\bar{a}) \notin I\}$$

It is clear from the construction of $P'$ that $P(I) = P'(I \cup I^c)$. Let $J = I \cup \{f\}$, recall $f = S(c_1, \ldots, c_n)$. Since there is a $c_i \notin adom(I)$, we have $I^c \subseteq J^c$. But now, $Q(I) \not\subseteq Q(J)$ contradicts $P'$ being monotone since: $Q(I) = P'(I \cup I^c)$ and $Q(J) = P'(J \cup J^c) = P'(I \cup I^c \cup J \cup J^c)$. □

**Remark 3.3** Note that this lemma can be used in very short proofs to show queries are not in $\mathcal{SP}$ (similar to how the Pumping lemma can be used for regular languages). For example, here is an elegant separation of semi-positive Datalog and (stratified) Datalog$^\neg$.

Consider the query $Q$ checking if there is a path of length one but not length two in a graph given by the edge relation e. Clearly, $Q$ is in stratified Datalog$^\neg$:

p$_1$ ← e(x,y). p$_2$ ← e(x,y),e(y,z). answer ← p$_1$,¬p$_2$.

But $Q$ cannot be in $\mathcal{SP}$ because for $I = \{$e(a,b)$\}$ and the fact e(b,c) with the constant c $\notin adom(I)$, we have $Q(I) \not\subseteq Q(I \cup \{$e(b,c)$\})$ witnessed by (the) answer.

A monotone query has a non-shrinking output for any fact $f$ that is added to the input. An adom-monotone query has non-shrinking output only for facts that each contain *at least one new* constant. As we will see, weak-adom-monotone queries have non-shrinking output for non-nullary facts that contain only new constants.

**Definition 3.4 (Weak-Adom-Monotonicity)** Let $I$ be a database instance over a schema $\sigma$, and $R$ a relation symbol in $\sigma$ of arity $n \geq 1$. A query is *weak-adom-monotone* if $Q(I) \subseteq Q(I \cup \{t\})$ for all inputs $I$ and facts $f = R(c_1, \ldots, c_n)$ that contain only constants $c_i$ that do not occur in $I$, i.e., $c_i \notin adom(I)$ for all $i$. We denote the class of weak-adom-monotone queries as $\mathcal{M}_{weak-adom}$.

It is easy to see that a query is weak-adom-monotone iff $Q(I) \subseteq Q(I \cup I')$ for all database instances $I$ and $I'$ with disjoint $adom(I)$ and $adom(I')$ and $I'$ not containing any nullary facts. With these Definitions in place, we can now present our main theorem:

**Theorem 3.5**

$$\mathcal{F}[\mathcal{N}_1] = \mathcal{M}_{adom} \quad \text{and} \quad \mathcal{F}[\mathcal{N}_2] = \mathcal{M}_{weak-adom}$$

Together with the results from [1, 3] and Lemma 3.2, this implies the following relation-ship between the defined query classes. Again, denote with $\mathcal{M}$ the set of monotone queries, and with $\mathcal{C}$ the set of computable queries.

**Corollary 3.6**

$$\begin{array}{ccccccc}
F[\mathcal{N}_0] & \subsetneq & \mathcal{F}[\mathcal{N}_1] & \subsetneq & \mathcal{F}[\mathcal{N}_2] & \subsetneq & F[\mathcal{N}_3] \\
\| & & \| & & \| & & \| \\
\mathcal{M} & \subsetneq & \mathcal{M}_{adom} & \subsetneq & \mathcal{M}_{weak-adom} & \subsetneq & \mathcal{C}
\end{array}$$

The rest of the section is devoted to proving the main theorem.

### 3.1 Proof of Main Theorem

**Lemma 3.7** $\mathcal{F}[\mathcal{N}_1] \subseteq \mathcal{M}_{adom}$

PROOF. We can generalize the proof idea for Lemma 6.6 in [3]. Let $Q$ be a query with $Q \notin \mathcal{M}_{adom}$ over the schema $\sigma$, and let $S$ be an $n$-ary relation symbol in $\sigma$. Assume towards a contradiction there is a coordination-free $\mathcal{N}_1$ transducer $\mathcal{T}$ computing $Q$. Since $Q \notin \mathcal{M}_{adom}$, we have a fact $w$ with $w \in Q(I)$ and $w \notin Q(I \cup \{f\})$ for a database instance $I$ and a fact $f = S(c_1, \ldots, c_n)$ with $c_i \notin adom(I)$ for some $i$. Since $\mathcal{T}$ is coordination-free, it computes the correct result $Q(I)$ on all networks $N$ and a Partitioning policy $\mathcal{P}$ with only performing $k$ heartbeat transitions for some $k \in \mathbb{N}$. Scenario (1): choose the network that contains only a single node $n_0$. We observe that this implies $\mathcal{P} \equiv \{n_0\}$. Now, scenario (2) consider input $I \cup \{f\}$ on the network $N' = \{n_0, n_1\}$. We choose a partitioning policy $\mathcal{P}'$ that assigns all ground atoms except $f$ to $n_0$; and $\mathcal{P}(f) = \{n_1\}$. Then, the content of the memory and system relations on $n_0$ are indistinguishable from the ones in scenario (1) since the transducer cannot possibly query the partitioning policy about $f$ since $f$ contains a constant $c_i$ that is not known to $n_0$. Thus, when the transducer at node $n_0$ is stepped $k$ times with only heartbeat transitions, it will output $w \in Q(I)$, among possibly other ground facts. However, this is a contradiction since $w \notin Q(I \cup \{f\})$. □

**Lemma 3.8** $\mathcal{M}_{adom} \subseteq \mathcal{F}[\mathcal{N}_1]$

PROOF SKETCH. Constructive. Given a computable query $Q \in \mathcal{M}_{adom}$, we construct a type $\mathcal{N}_1$ coordination-free transducer $\mathcal{T}$ that distributedly computes $Q$. The proof is based on the basic idea of the proof for Lemma 6.3 in [3]. We use the same transducer, but *emulate* adom via a memory relation. The emulation is not perfect: adom always is an underestimate of the constants in $adom(I)$; however, every fair run of a transducer network will eventually reach a state in which adom equals $adom(I)$. Initially, adom is computed from the local active domain. Then, for each $k$-ary input relation $R$ of $Q$, we add the following broadcast rules:

```
adom@all(X) ← R(X,_,...,_).
adom@all(X) ← R(_,X,...,_).
...
adom@all(X) ← R(_,_,...,X).
```

Like with the transducer $\mathcal{T}$ in the proof for Lemma 6.3 in [3], we compute $Q$ over the local state and output its result whenever ready is true. In a fully connected network, eventually, the complete active domain is known to every node and thus the complete query result is output. Clearly, when all data is partitioned on one node $n_0$, then $n_0$ will already output the whole query result even when stepped only with heartbeat transitions.

Let the input to the transducer be the database instance $I$. We now just have to show that whenever during the process of computing $Q(I)$ a fact $w$ is output, $w \in Q(I)$. Tuples are output only if ready is true. That is at a point when for adom $\subseteq adom(I)$, the transducer has both positive and negative information about all facts $R(\bar{a})$ with $R \in \sigma(I)$ with $\alpha(R) = k$, and $\bar{a} \in \text{adom}^k$; let us call the instance that is in the transducer's memory when ready is true $I'$. Let $I' \cup J = I$ for some minimal $J$. If $J = \emptyset$, we are immediately done since by definition $\mathcal{T}$ computes the correct result. Let $J \neq \emptyset$. It is now easy to see that all facts $f \in J$ must contain a constant $c \notin adom(I')$ (otherwise $\mathcal{T}$ would know about $f$). But now, a simple inductive argument over the definition of adom-monotone shows $Q(I') \subseteq Q(I' \cup J) = Q(I)$, and thus the output fact $w$ is in $Q(I)$. □

**Lemma 3.9** $\mathcal{F}[\mathcal{N}_2] \subseteq \mathcal{M}_{weak-adom}$

PROOF SKETCH. The proof is very similar to the proof of Lemma 3.7; with the difference that in addition to $f = S(c_1, \ldots, c_n)$ having a constant $c_i \notin adom(I)$, we also require that none of $c_i$ is in $adom(I)$. This restriction then allows to place $I$ and $f$ on two different nodes by an partitioning policy $\mathcal{P}'$ that is compatible with $\mathcal{N}_2$. The obtained contradiction is the same: when $I$ is placed in a one-node network, the coordination-free transducer outputs $w$ with only heartbeat transitions, but when $I$ and $f$ is provided on two distinct machines, the node with $I$ cannot tell the difference and also outputs $w$—which is not part of $Q(I \cup \{f\})$. □

**Lemma 3.10** $\mathcal{M}_{weak-adom} \subseteq \mathcal{F}[\mathcal{N}_2]$

PROOF SKETCH. The relational transducer behaves similar to the transducers for monotone and adom-monotone queries. Each transducer collects data from itself and its neighbors including knowledge about the active domain. The general idea is to allow each node to answer the following question: "For the relation $R$, an attribute index $i \in \{1, \ldots, \alpha(R)\}$ and a constant $a \in$ adom: *Do I have all facts $R(\bar{x}, a, \bar{x}') \in I$ from the global input?* If the query is answered affirmative for all $R \in edb(Q)$, all column indexes $i$, and all $a$ contained in the current adom, then it is safe to run $Q$ on the local input even though adom might not be complete, i.e., adom $\subsetneq adom(I)$. The reason being that possibly missing facts do not use any constant $a$ in adom and thus the computed output is a fine subset of the global output due to the weak adom monotonicity of $Q$.

A coordination-free type $\mathcal{N}_2$ transducer with system relations Id, All, and Local$_R$ for each edb relation $R$ would implement a broad-cast with acknowledgements of single facts, and when all facts for a relation $R$, a column-index $i$ and a domain-constant $a$ have been acknowledged from a node $n_1$, then the node $n_1$ can be sent a tuple indicating that it now can affirmatively answer the question for $R, i, a$.

Since we again, implement broadcast to also be directly fed to the sending node, it is clear that when all data is partitioned to one node that the algorithm will detect completeness for its known active domain and thus the correct result is computed even with only heartbeat transitions. □

## 4. RELATED WORK AND CONCLUSION

This work was inspired by conjectures of Joe Hellerstein in his PODS 2010 keynote [2]. The CALM conjecture suggests a close connection between <u>C</u>onsistency <u>a</u>nd <u>L</u>ogical <u>M</u>onotonicity. As shown in [1], monotone queries can be implemented in a coordination-free distributed manner while non-monotone programs cannot. We showed that when we allow the nodes of the distributed system to draw conclusions about the *non-existence* of facts, then seemingly non-monotone reasoning can still be performed coordination-free in a distributed manner. The degree of information we provide to individual nodes determines how much the monotonicity requirement of the queries can be weakened. The contribution of this work is to characterize the coordination-free languages $\mathcal{F}[\mathcal{N}_1]$ and $\mathcal{F}[\mathcal{N}_2]$ introduced in [3] exactly by means of two novel notions of "weakened" monotonicity. Together with the results in [1] and [3], this yields the following characterization:

$$\begin{array}{ccccccc}
F[\mathcal{N}_0] & \subsetneq & \mathcal{F}[\mathcal{N}_1] & \subsetneq & \mathcal{F}[\mathcal{N}_2] & \subsetneq & F[\mathcal{N}_3] \\
\| & & \| & & \| & & \| \\
\mathcal{M} & \subsetneq & \mathcal{M}_{adom} & \subsetneq & \mathcal{M}_{weak-adom} & \subsetneq & \mathcal{C}
\end{array}$$

**Future Work.** Recall that the algorithms used in our constructive proofs postponed executing the query $Q$ until the locally available data seemed complete. For type $\mathcal{N}_2$ networks, this means that sub-problems that have disjoint active domains are solved independently from each other. Note that the algorithm for solving win-move as presented in [3] *incrementally* produces *partial* results at a finer granularity. Finding robust notions and cost-models that consider various degrees of coordination is an interesting avenue for future work.